\documentclass[review,number,sort&compress]{elsarticle}
\journal{Nuclear Instrumentation and Methods Section A}
\usepackage{graphicx} \usepackage{amsmath} \usepackage{fullpage}
\usepackage{color} \usepackage{physics}
\usepackage{tikz} \usepackage{ amssymb } \usepackage{float}
\usepackage{courier} \usepackage{soul} \usepackage{threeparttable}
\usepackage{ mathrsfs } \usepackage{bm}
\newcommand{\atom}[2]{$^{#1}$#2}
\newcommand{\oxy}[0]{\atom{18}{O}(p,$\gamma$)\atom{19}{F}}
\newcommand{\mg}[0]{\atom{25}{Mg}(p,$\gamma$)\atom{26}{Al}}

\newcommand{\flourine}[0]{\atom{19}{F}(p,$\alpha_2$$\gamma$)\atom{16}{O}}
\newcommand{\carbon}[0]{\atom{12}{C}(p,$\gamma$)\atom{13}{N}}

\definecolor{lightblue}{rgb}{.90,.95,1}
\sethlcolor{lightblue}

\begin{document} 
\begin{frontmatter}
\title{$\gamma$-ray Spectroscopy using a Binned Likelihood Approach}
\author[unc,tunl]{J. R.~Dermigny\corref{cor1}} 
\author[unc,tunl]{C.~Iliadis} 
\author[unc,tunl]{M. Q.~Buckner \corref{cor2}} 
\author[unc,tunl]{K. J.~Kelly} 
\cortext[cor1]{Corresponding author}
\cortext[cor2]{Present address: Lawrence Livermore National Laboratory, 7000 East Avenue, Livermore, CA 94550}
\address[unc]{The University of North Carolina at Chapel Hill, Chapel Hill, North Carolina 27599-3255, USA}
\address[tunl]{Triangle Universities Nuclear Laboratory, Durham, North Carolina 27708-0308, USA}
\begin{abstract} 
The measurement of a reaction cross section from a pulse height spectrum is a ubiquitous problem in experimental nuclear physics. In $\gamma$-ray spectroscopy, this is accomplished frequently by measuring the intensity of full-energy primary transition peaks and correcting the intensities for experimental artifacts, such as detection efficiencies and angular correlations. Implicit in this procedure is the assumption that full-energy peaks do not overlap with any secondary peaks, escape peaks, or environmental backgrounds. However,  for complex $\gamma$-ray cascades, this is often not the case. Furthermore, this technique is difficult to adapt for coincidence spectroscopy, where intensities depend not only on the detection efficiency, but also the detailed decay scheme. We present a method that incorporates the intensities of the entire spectrum (e.g., primary and secondary transition peaks, escape peaks, Compton continua, etc.) into a statistical model, where the transition intensities and branching ratios can be determined using Bayesian statistical inference. This new method provides an elegant solution to the difficulties associated with analyzing coincidence spectra. We describe it in detail and examine its efficacy in the analysis of \oxy{} and \mg{} resonance data. For the \oxy{} reaction, the measured branching ratios improve upon the literature values, with a factor of $4$ reduction in the uncertainties. 
\end{abstract} 
\begin{keyword}
    $\gamma$-ray spectroscopy \sep GEANT4  \sep binned likelihood \sep Bayesian analysis
\end{keyword}
\end{frontmatter}
\linespread{1.3}
\section{Introduction} \label{sec:introduction}
Of main interest in the study of nuclear capture reactions are (i) the fraction of primary $\gamma$-ray decays from the compound state to lower-lying levels, i.e., the primary $\gamma$-ray branching ratios, and (ii) the total number of nuclear reactions that took place. Traditionally, the net intensities of all full-energy primary transition peaks are measured and are carefully corrected for the detection efficiency to determine the reaction yield. While this is an attractive option for simple spectra, $\gamma$-ray cascades are often sufficiently complex that this ``peak-by-peak" analysis is challenging. Coincidence summing effects, coupled with angular correlations, complicate the analysis by requiring cumbersome corrections to each measured peak. 
\par
The situation worsens for $\gamma\gamma$-coincidence spectroscopy. A coincidence spectrometer operates by requiring multiple hits across two or more detectors. By applying timing or energy conditions (or \textit{cuts}), unwanted signals, such as environmental backgrounds, can be minimized; this affords an increase in detection sensitivity. Each coincidence event satisfies timing and energy gates; the measured peak not only depends on the detection efficiencies, but also on the detailed decay of the entire $\gamma\gamma$-cascade initiated by a primary transition. This effect, compounded by the challenges described above, makes the analysis of coincidence spectra difficult.
\par
In light of the above challenges, we present a new method of spectral analysis that has two innovations. First, we model our data using a binned likelihood function. This allows us to fit the entire spectrum --- every full-energy peak, as well as their Compton distributions and escape peaks --- using Monte Carlo simulated spectra, or \textit{templates}. Second, we determine the fraction of the experimental spectrum belonging to each template using a Bayesian statistical approach. 
This allows for the extraction of the primary $\gamma$-ray branching ratios and the total number of reactions, not \textit{only} from individual full-energy peaks, but from the entire pulse height spectrum. Explicit corrections for coincidence summing and angular correlations are no longer necessary, as these effects are implicitly included in the Monte Carlo simulations used to generate the templates. This method applies to both singles and coincidence spectra, removing many difficulties faced in a traditional ``peak-by-peak" analysis.
\par
The first part of this method, i.e., use of the likelihood function to study capture reactions, was first applied to high-purity germanium and coincidence spectra during the analysis of \atom{17}{O}(p,$\gamma$)\atom{18}{F} reaction data \cite{buckner15}. We present here a more complete description of this new methodology. In Section~\ref{sec:analysis}, we define the statistical model used in our analysis.
The method for generating templates is discussed in Section~\ref{sec:mc}. The experimental apparatus is described in Section~\ref{sec:equipment}. In Sections~\ref{sec:18o} and \ref{sec:25mg}, we demonstrate the method by measuring the resonance strengths of two well-known resonances: the E$_r^{\text{lab}}=150.5 \pm 0.5$ keV \cite{becker} resonance in \oxy{} and the E$_r^\text{lab}=316.7 \pm 0.5$ keV \cite{endt90} resonance in \mg{}. A summary and conclusions are presented in Section~{\ref{sec:conclusion}. Lastly, an appendix is provided to motivate the use of Bayesian inference.
\par
\section{Analysis Method} \label{sec:analysis}
A binned pulse height spectrum, where the pulse heights correspond to energy dispersion in a detector, consists of contributions arising from different sources: room background, beam-induced background, and the reaction of interest, itself consisting of primary $\gamma$-ray transitions and their corresponding secondary decays. Each of these contributions requires a template, containing the entire pulse height distribution (e.g., full-energy peaks, Compton continua, and escape peaks) unique to that source. We discuss strategies for generating the templates in Sec.~\ref{sec:mc}. In the following, we define a formalism where, for each template, $j$, we predict the fraction of events, $F_j$, present in the experimental spectrum that results from source $j$.
Using this formalism, we can calculate the number of reactions (or decays) attributable to each source, thus furnishing the reaction intensity as well as the branching ratios.
\par
To model our data, we adopt the \textit{extended} binned likelihood function \cite{barlow90}. 
This likelihood function, that is, the probability of obtaining the data, $\bm{D}$, given the $m$ template fractions, $\bm{F}$, is given by \cite{barlow}:
\begin{equation} \label{eq:likelihood}
P( \bm{D} | \bm{F}) = \bigg[ \sum_{i=1}^{n} D_i \ln f_i - f_i \bigg] + \bigg[\sum_{i=1}^{n} \sum_{j=1}^{m} a_{ji} \ln
A_{ji} - A_{ji} \bigg],
\end{equation} 
where, for each of the $n$ bins, $i$, $f_i$ is the total number of events contributed by all the templates, and $A_{ji}$ and $a_{ji}$ are the predicted mean and observed number of events in template $j$, respectively. The $A_{ji}$ account for the statistical fluctuations in the $a_{ji}$, as they are sampled from finite Monte Carlo calculations.
The $f_i$ are given by:
\begin{equation} \label{eq:fsubi}
f_i = \sum_{j=1}^{m} \frac{A^{data}}{ A_j^{sim} } F_j A_{ji}\ ,
\end{equation}
where $A^{data}$ and $A_j^{sim}$ are the total areas (within the fitted region) of the measured spectrum and of template $j$, respectively.
In a previous study \cite{buckner15}, the estimates for the fractions, $F_j$, were obtained through maximization of the likelihood using the \textsc{Minuit} library \cite{minuit}. In the present work, however, we have decided to follow an entirely different approach.
Instead, we apply Bayes' theorem \cite{bayes} to build a full probability model for our data and parameters. This allows some practical advantages over a maximum-likelihood estimate. Most importantly, using a Bayesian data analysis we can derive probability density functions for the parameters, e.g., the fraction values or source intensities. These probability distributions can then be used to calculate uncertainties and place meaningful upper-limits on weak transitions. A brief review of Bayesian statistical inference is given in the appendix to motivate its use in the present analysis.
\par
In a Bayesian framework, we make inferences using the multivariate joint posterior distribution, $P( \bm F | \bm D)$, as defined by Bayes' theorem:
\begin{equation}
\label{eq:bayes}
  P( \bm F | \bm D) = \frac{ P( \bm D |  \bm F )P( \bm F )    }{\int_{ \bm F}P(\bm D |\bm F)P( \bm F)} \;\;\;,
\end{equation}
where $P( \bm F)$ is the joint prior probability function for the model parameters and $P( \bm D | \bm F)$ is the likelihood function, as defined by Eq.~\ref{eq:likelihood}.
In constructing the joint prior distribution, we assume that each fraction value has a prior distribution which is independent from the rest. Further, for each parameter, we adopt a scale-invariant, non-informative Jeffreys prior \cite{jeffreys}. This choice was motivated by the requirement that the prior convey equal probability per decade, i.e., that each fraction value is as likely to be in range (0.001,0.01) as in (0.1,1.0). Thus, the joint prior distribution is given by
\begin{equation}
P(\bm{F}) = \prod_{j=0}^{m} \bigg[ \frac{1}{F_j}\bigg] \;\;\;.
\end{equation} 
The joint posterior distribution is then calculated using the \textit{evidence} procedure \cite{MacKay}, where the $A_{ji}$ are replaced by their maximum-likelihood estimates using the method of Barlow \textit{et al.} \cite{barlow}. This adjustment eliminates the $A_{ji}$ as nuisance parameters, thereby making the analysis more tractable.
\par
The problem of finding the contribution of each source has so far been formulated in terms of the fraction values, $F_j$. However, in the study of nuclear reactions, the number of decays that took place is the more interesting observable. Posterior distributions describing the individual source intensities, as well as the total intensity, are obtained in two steps.
First, a Markov chain Monte Carlo routine is used to sample the $F_j$ from the joint posterior distribution.  
This routine uses the \textit{Metropolis-Hastings} algorithm, with a multivariate Gaussian as the proposal distribution. The Markov chain is run for $2.5 \times 10^6$ iterations. Of these, $5 \times 10^4$ ($2\%$) are considered \textit{burn-in} and discarded. The remaining iterations are then thinned, keeping only every $50^{th}$ sample, and then checked for convergence via the Heidelberg and Welch diagnostic.
Then, the sampled fraction values, $F_j$, are scaled to obtain the number of events of source $j$, $N_j$, that gave rise to the measured spectrum \cite{buckner15}
\begin{equation}  \label{eq:partial}
N_{j} = \frac{ A^{data}}{ A_j^{sim}}  F_j N_j^{sim},
\end{equation}
where $N_j^{sim}$ is the number of simulated compound nucleus decays used to generate template $j$. For source contributions which describe transitions in a reaction of interest (as opposed to background), the total intensity, $N_R$, can be calculated at each iteration by summing the relevant $N_j$.
The samples are then histogrammed to produce posterior probability distributions for the observed transition intensities, $P(N_j|\bm D)$, and the total, $P(N_R|\bm D)$. For posterior distributions corresponding to weak or unobserved transitions (i.e., with a significant probability density at zero) we report an upper-limit on the transition intensity using the $95\%$ credible interval (See Appendix).
The posterior distributions describing observed transitions, as well as the total intensity, are found to be symmetric and unimodal. 
For these, we estimate their intensity using the mean of the distribution, where the uncertainty corresponds to the region defined by the 68\% highest posterior density interval.
\par
For our purposes, each template describing a component of the reaction of interest represents a primary $\gamma$-ray transition from a resonant state; the $\langle N_{j} \rangle$, obtained from the $P(N_j|\bm D)$, are then the partial number of primary decays of component $j$ that contributed to the experimental spectrum, while $\langle N_R \rangle$ is the total number of reactions, obtained from $P(N_R|\bm D)$.
The branching ratios for each primary transition follows from the partial and total number of reactions,
\begin{equation} \label{eq:br}
B_j = \frac{\langle N_j\rangle }{\langle N_R \rangle}.
\end{equation} 
The $\gamma$-ray branching ratios and total number of reactions are therefore obtained without explicitly correcting each individual full-energy peak for detector efficiencies, angular correlation effects, or coincidence summing, as is required in traditional $\gamma$-ray spectroscopy. This is made possible by incorporating these experimental artifacts implicitly into the simulated template spectra. In the next section, we will discuss these points in detail.

\section{Geant4 Simulations}
\label{sec:mc}
\begin{figure}
    \centering
    \includegraphics[width=.40\textwidth]{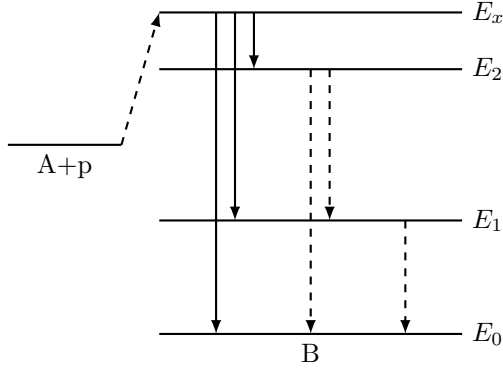}
    \caption{A generic proton capture with decay scheme. For illustrative purposes we show the compound state, E$_x$, which can decay to the ground state, E$_0$, or to the first two excited states (E$_1$, E$_2$). These decays are called \textit{primary} transitions (solid arrows). Each of these, except for the ground state transition, E$_x \rightarrow$ E$_0$, gives rise to \textit{secondary} transitions, e.g., E$_2 \rightarrow$ E$_0$,  E$_2$ $\rightarrow$ E$_1$ $\rightarrow$ E$_0$ (dashed lines). For this example, three separate templates would be created, each corresponding to a possible primary transmission, that would be fitted to the experimental spectrum. }
    \label{fig:genreaction}
\end{figure}
\subsection{General Strategy}
\label{sec:strategy}
The form of the likelihood function (Eq. \ref{eq:likelihood}) is predicated on the presumption that the experimental spectrum can be described by a set of templates. Therefore, it is important to identify the source of every peak, e.g., primary or secondary transition, environmental background, or beam-induced background. Based on the results of this analysis, templates must be generated for each source.  We will discuss the procedure for environmental and beam-induced background templates in Sec.~\ref{sec:bkgd}. For the reaction of interest, it is important that (i) every observed primary $\gamma$-ray transition is described by a template, and (ii) each template represents the detector response to the initial (primary) decay \textit{and} all subsequent secondary decays. 
An example decay scheme is shown in Fig.~\ref{fig:genreaction}. The target nucleus, \textit{A}, captures a proton. The resulting compound nucleus, \textit{B}, deexcites from the resonant level E$_x$ to a lower lying level: E$_\text{2}$, E$_\text{1}$, or the ground state, E$_\text{0}$. Each of these deexcitations represents a \textit{primary} transition and thus requires a simulated template. The E$_x$ $\rightarrow$ E$_\text{2}$ and E$_x$ $\rightarrow$ E$_\text{1}$ transitions are accompanied by \textit{secondary} $\gamma$-ray decays, (shown as dashed lines). These transitions must also be included in the primary transition templates.
A code was written which uses the G{\sc eant}4.9.6 \cite{geant4,geant2} framework to simulate the compound nucleus deexcitation $\gamma$-cascade during a nuclear reaction.
\par
The simulated $\gamma$-ray cascades are used to populate template histograms. The cascade begins at the energy of the compound nucleus in the resonant state, E$_x$. The compound level decays to the secondary state (e.g., E$_2$), emitting a unique $\gamma$-ray of energy E$_x$-E$_2$. The subsequent secondary decays are then simulated by randomly sampling over the known secondary $\gamma$-ray decay branchings. The simulation continues until the $\gamma$-cascade terminates at the ground state, E$_0$. All simulated $\gamma$-rays originate from the ion beamspot on the target and are tracked as they interact with the spectrometer and the  environment (e.g., the beamline, target holder, target backing, shielding, cooling water, etc.) via photoelectric absorption, Compton scattering and pair production. Energy deposition in the active detector volume is recorded for each event and written to an output file. The output, accumulated over many simulated decays, is then used to construct a template histogram with the same energy and timing gates applied to the measured pulse height spectrum.
\par
\subsection{Corrections to Simulated Templates} \label{sec:corrections}
A number of corrections must be performed before simulated templates can be used to analyze a measured pulse-height spectrum.  For instance, simulating a $\gamma$-ray cascade requires input of all excitation energies and secondary branching ratios. While this is sufficient in reproducing the approximate \textit{location} of many full-energy peaks at their observed energies, the $\gamma$-ray energies of primary decays may need to be adjusted in the simulations by several keV to account for Doppler shifts.
\par
It is also important that the simulation reproduce the measured peak \textit{widths}. This is achieved in two steps. First, we convolve each raw simulated spectrum with a Gaussian of width $\sigma(E)$, where $\sigma(E)$ is the energy-dependent detector resolution function, characterized by the measured full width at half maximum (FWHM) of room background or secondary-transition full-energy peaks. The width, $\sigma$, of each peak is then determined using the relationship $\text{FWHM}=2\sqrt{2\ln2}\sigma$. Second, we may need to consider an additional broadening for primary transition peaks. For example, $\gamma$-ray transitions from a short-lived resonant state are frequently Doppler broadened. Also, the observed peak width may reflect the target thickness if the reaction proceeds via a direct (i.e., non-resonant) mechanism. In such cases, the simulated full-energy peaks are broadened until they match the widths of their experimental counterparts. 
\par
Depending on the experimental alignment and the angular momentum coupling involved in the reaction of interest,
the $\gamma$-ray emission following a nuclear reaction may be anisotropic. 
For these transitions, the angular distribution for the emitted radiation is described by \cite{iliadis}:
\begin{equation}
 \label{eq:W}
     W_{ij}(\theta) = 1 + a_{2}P_2(\cos\theta) + a_{4}P_4(\cos\theta) 
 \end{equation}
where $a_{2}$ and $a_{4}$ are angular correlation coefficients that depend on the angular momentum coupling involved in the reaction, $\theta$ is the polar coordinate with the z-axis oriented toward the direction of the incident beam or previous radiation, and $P_2$ and $P_4$ represent the second and fourth order Legendre polynomials. For a known angular correlation, the coefficients can be directly adopted in Eq.~\ref{eq:W}. If the angular correlation has not been measured yet, the coefficients can frequently be calculated (see Appendix D in Ref.~\cite{iliadis}). When simulating the decay of the compound nucleus, $\gamma$-rays are emitted according to an angular probability distribution, where the probability for emittance into the solid angle $\text{d}\Omega$ is weighted such that $p(\Omega)\text{d}\Omega=W(\theta)\textit{d}\Omega$. This correction is frequently only necessary for primary transition $\gamma$-rays, where the emission is correlated with the direction of the incident proton beam. Angular correlations for secondary transitions, with respect to either the incident beam or previously emitted $\gamma$-rays, have a much smaller effect on spectra.
\par
\subsection{Background Contaminants}
\label{sec:bkgd}
Radiation from radionuclides present in the environment, as well as from beam-induced reactions or cosmic-ray interactions, contribute unwanted background to experimental spectra. Environmental radionuclides, such as \atom{40}{K} and \atom{208}{Tl}, and secondary radiation from cosmic-rays can be accounted for by measuring the room background in the run geometry for an extended period of time. Since the background may vary with meteorological conditions, two background spectra should be recorded, one before and one after the reaction measurement. The combined pulse height spectrum can then used as a background template.
Beam-induced reactions result from contaminants (e.g., \atom{19}{F}, \atom{11}{B} and  \atom{12}{C}) in the target or backing. 
Their $\gamma$-ray contributions can be simulated in the same manner as the reaction of interest, or measured directly by an off-resonance run.

\subsection{Uncertainties}  
\label{sec:unc}

The exact form of the templates used for the different primary transitions is affected by many aspects of the simulation. For instance, the intensities of the secondary transition full-energy peaks depend on the branching ratios assumed for the $\gamma$-cascade. Further, since the Compton continua are considered in the fit, scattering off of the surrounding materials, e.g., the detector dead-layer, the shielding, and the NaI(Tl) annulus, influences the spectra. In the case of branching ratios, literature values are often measured to acceptable tolerances, so these effects should be small. In cases where strong secondary transitions have broad uncertainties (i.e., $\pm 5\%$), the sensitivity of the final results to these transitions should be explored.

To estimate the effect of the surrounding materials, a sensitivity study focused on the simulated detector dimensions should be performed. While this can be done in many different ways, it is particularly convenient to explore the sensitivity of the simulated total efficiency. This entails calculating the efficiency for an ensemble of simulated detector systems, all having a slightly different geometry. The variance in the calculated efficiencies then provides a direct measure of the systematic uncertainty introduced by the detector dimensions. This method can also demonstrate the accuracy of the simulation, since a comparison to an experimentally determined efficiency can easily be made. A discussion of systematic uncertainties inherent to our own simulation is presented in the next section.

\section{Equipment }
\label{sec:equipment}
\subsection{Ion beams and targets}
\label{sec:accelerators}
We applied the analysis method outlined above to resonance data collected at the Laboratory for Experimental Nuclear Astrophysics (LENA), in Durham, North Carolina. The LENA facility houses two accelerators. The JN Van de Graaff accelerator is capable of producing proton beams of up to I$_p$ = 100 $\mu$A on target in the energy range below E$_p^\text{lab}$=1 MeV. The bombarding energy of the JN accelerator was calibrated using direct-capture primary $\gamma$-rays from \carbon{}.
The second accelerator consists of a high-current, low-energy electron cyclotron resonance ion source (ECRIS). The LENA ECRIS produces a maximum beam current of  I$_p\approx 2.0$ mA on target, and it is used to collect data at bombarding energies below E$_p^{lab} =200$ keV. The JN and ECRIS were used to collect the \mg{} and \oxy{} resonance data, respectively. Detailed information regarding the accelerators can be found in Ref.~\cite{Cesaratto}. 
\par
The \atom{18}{O} targets were prepared by anodizing 0.5-mm-thick tantalum backings in \atom{18}{O}-enriched water. According to the manufacturer, the water composition (in atom $\%$) was 99.3 (\atom{18}{O}), 0.5 (\atom{16}{O}) and 0.2 (\atom{17}{O}). Such targets have a well-defined stoichiometry (Ta$_2$O$_5$) and are stable under high-intensity proton bombardment \cite{buckner15} . The \atom{25}{Mg} target was produced by thermal evaporation as in Ref.~\cite{mgO}: a mixed powder consisting of isotopically enriched MgO (99 atom $\%$ \atom{25}{Mg}) and Zr$_2$, the reducing agent,  was seated in a resistively heated tantalum boat, while evaporation onto a $0.5$-mm-thick tantalum backing was monitored \textit{in situ} via a thin film thickness monitor. Prior to deposition, the surfaces of the tantalum backings for both targets were etched and then outgassed through resistive heating in vacuum to reduce the presence of contaminants.

\subsection{Detectors}
\label{sec:detectors}
The detection system used at LENA is shown in Fig.~\ref{fig:detectors}. The $\gamma\gamma$-coincidence spectrometer features a 134$\%$ HPGe detector, oriented at $0^{\circ}$ with respect to the beam axis, surrounded by a 16-segment NaI(Tl) annulus. The detectors are surrounded on five sides by anti-coincidence plastic scintillator paddles. The use of this spectrometer for $\gamma\gamma$-coincidence spectroscopy has been reported previously in Refs. \cite{buckner15,cesaratto13,buckner12}. A more detailed description of the $\gamma\gamma$-coincidence techniques employed at LENA is given in Ref.~\cite{longland06}. 
\par 
\begin{figure}[h!]
    \centering
    \includegraphics[width=.45\textwidth]{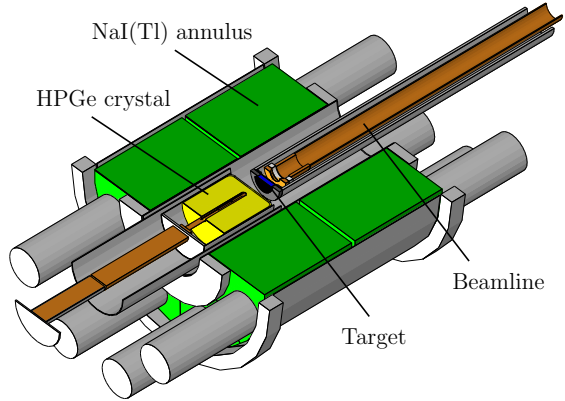}
    \caption{(Color Online) The HPGe crystal (yellow) is located in close geometry to the target. Both the target and the HPGe detector are surrounded by a 16-segment NaI(Tl) annulus (green). } 
    \label{fig:detectors}
\end{figure} \par
Energy and timing signals were processed using standard NIM and VME modules. Events were sorted off-line using the acquisition software JAM \cite{JAM}, and all coincidence energy and timing gates were set in software. The HPGe-NaI(Tl) coincidence timing gates were set sufficiently wide ($\approx 500$ ns)  because of the long mean lifetime of the $E_x = 197$ keV level ($\tau  = 128.8 \pm 1.5 $ ns \cite{Tilley}) in \atom{19}{F}. 
Counting rates were kept low ($< 1000$ cps) to reduce the effects of chance coincidences by the random-summing of uncorrelated events.
In a two-dimensional NaI(Tl) energy versus HPGe energy histogram, wide trapezoidal gates were drawn with the condition that the summed energy, $ \text{E}_{\gamma}^{\text{HPGe}} + \text{E}_{\gamma}^{\text{NaI(Tl)}} $, falls within user-defined high and low energy thresholds. The low-energy threshold is employed to reduce room-background contributions, such as those due to \atom{40}{K} and \atom{208}{Tl}, and thus reduces the cosmic ray muon-induced background. The high energy threshold is chosen to exceed the $Q$-value of the reaction plus the center-of-mass kinetic energy. For the \oxy{} data ($Q = 7993.5994 \pm 0.0011$ keV \cite{wang}), the summed energy condition was
\begin{equation} \label{eq:1}
3.5 \text{  MeV} \le \text{E}_{\gamma}^{\text{HPGe}}+ \text{E}_{\gamma}^{\text{NaI(Tl)}} \le 8.5 \text{  MeV},
\end{equation}
and for the \mg{} data ($Q = 6306.31 \pm 0.05$ keV \cite{wang}) we employed the values of
\begin{equation} \label{eq:2}
3.5 \text{  MeV} \le \text{E}_{\gamma}^{\text{HPGe}}+ \text{E}_{\gamma}^{\text{NaI(Tl)}} \le 7.0 \text{  MeV}.
\end{equation}
\par
To generate reliable templates, an accurate model of the spectrometer is of paramount importance. To that end, the HPGe and NaI(Tl) detectors have been modeled extensively using the GEANT4 toolkit. A previous study \cite{carson2010} used computed tomography to precisely determine the internal geometry of the HPGe detector. These dimensions were then incorporated into a simulation, where it was shown that the simulated relative peak efficiency (of $\eta^P_{4.44\text{ MeV}}$ to $\eta^P_{11.66 \text{ MeV}}$) was in agreement with the experimentally determined value within an uncertainty of $1.6\%$, demonstrating that the energy dependence of the detector efficiency was well modeled.
The dimensions of the surrounding NaI(Tl) annulus (provided by the manufacturer) were then incorporated into the detector simulation by Howard \textit{et al.}~\cite{choward}, allowing a detailed analysis of the full $\gamma\gamma$-coincidence spectrometer.
In the same study, a spectrum was taken from a calibrated \atom{22}{Na} source and compared to a simulated singles spectrum, normalized to the equivalent number of decays. The simulated and measured intensities of the full-energy $1275$-keV peak were in agreement, to within $2\%$ error, suggesting that the simulated HPGe detector response is accurate. 
The simulated NaI(Tl) response was studied by applying a coincidence gate with the condition that the NaI(Tl) annulus fully detects two 511-keV $\gamma$-rays emitted during the decay. The simulated and measured (coincidence) intensity of the $1275$-keV peak were again in agreement, to within $3\%$ error. These tests demonstrate that the simulated spectrometer faithfully reproduces singles and coincidence spectra above a low-energy limit (determined by the electronic pulse height threshold). For the present work, the detector dimensions reported in Ref.~\cite{carson2010} (HPGe) and Ref.~\cite{choward} (NaI(Tl) annulus) were adopted for all simulations. 
\par
Sensitivity tests have been conducted to determine systematic effects inherent to the simulated spectrometer. Howard \textit{et al.}, following the procedure suggested in Section~\ref{sec:unc}, determined that the uncertainties in the detector geometry (e.g., HPGe  dead-layer thickness, NaI(Tl) crystal length) amounted to a systematic uncertainty of only 1.3\% in the simulated total efficiency. Further, in Refs.~\cite{buckner15,mqbthesis}, Buckner \textit{et al.} compared a simulated full-energy peak efficiency curve with experimentally determined efficiencies from  $^{60}$Co, $^{14}$N($p,\gamma$)$^{15}$O, $^{18}$O($p,\gamma$)$^{19}$F, and $^{27}$Al($p,\gamma$)$^{28}$Si reaction data. The simulated curve was found to differ from the fitted efficiencies by 5\%. A similar test performed on $^{14}$N($p,\gamma$)$^{15}$O coincidence data found that this result holds for gated spectra as well. Therefore, in the present analysis, we adopt a systematic uncertainty of $5$\% for the simulated detector response.

\section{$^{18}$O(p,$\gamma$)$^{19}$F} \label{sec:18o}
\begin{figure*}
    \includegraphics[width=\textwidth]{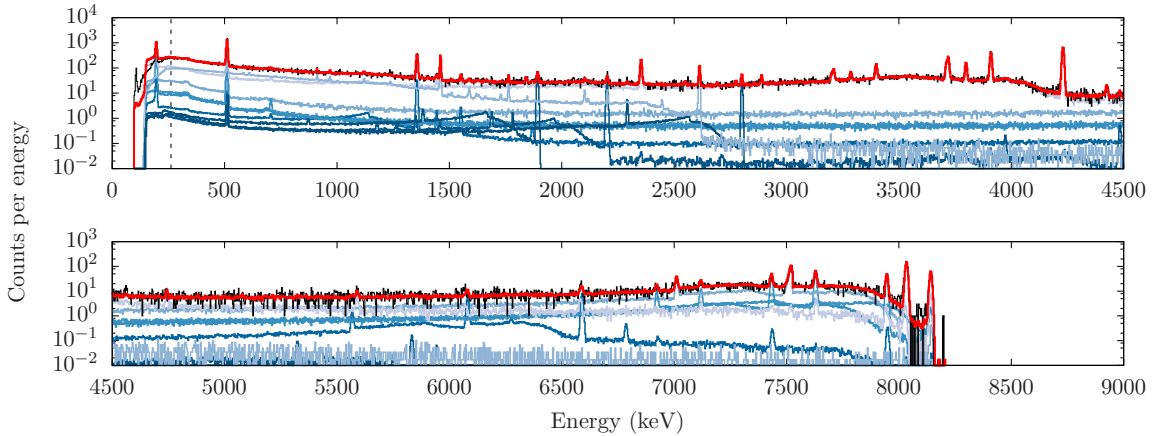}
    \caption{(Color online) The measured pulse height spectrum (black) obtained for the E$_r^{lab}$ = 151 keV resonance in \oxy{}. The fit limits are shown as black dashed vertical lines. The contributions of the different primary decays (i.e., scaled templates), including all associated secondary transitions, are shown in shades of blue/purple. The red line represents the sum of all the templates and so should match the experimental spectrum (black) over the region included in the fit. Below the lower fit limit ($\approx 200$ keV) these two lines diverge near the electronic pulse height threshold set in the experiment.}
    \label{fig:pulseheightspectrum}
\end{figure*} 
We will first apply our new technique to the $151$-keV resonance in \oxy{}, which has a well-known resonance strength. Additionally, it has a simple decay scheme, consisting of just seven primary $\gamma$-ray decays.
The ECRIS (Sec.~\ref{sec:accelerators}) was used to accumulate a charge of 38 mC, with an average beam intensity of $\approx 30$ $\mu$A on target. The bombarding energy of the H$^+$ beam was E$_p^{lab} = 155 $ keV, corresponding to the plateau of the thick-target excitation function. The measured resonance data were sorted into singles and coincidence spectra, with the coincidence energy gates defined by Eq.~\ref{eq:1}.
\par
The primary $\gamma$-ray decay scheme for the 151-keV resonance in the \oxy{} reaction, according to Ref.~\cite{Tilley}, is shown in Fig.~\ref{fig:oxydecayscheme}.  
Each of the seven primary $\gamma$-ray transitions is shown, with the strongest transitions proceeding to the 110-keV and 3908-keV excited states. All primary $\gamma$-ray transitions were observed in our experimental spectrum and thus required a template in the analysis.  The reaction templates were generated using the procedure described in Sec.~\ref{sec:strategy}, where the secondary $\gamma$-ray branching ratios were adopted from Ref.~\cite{Tilley}. The $\text{E}_p^{lab}=151$ keV resonance has spin-parity of $J^{\pi}=\frac{1}{2}^+$ \cite{wiescher}. Thus, in an experiment with an unpolarized beam and unpolarized target, the magnetic substates are equally populated. This gives rise to an isotropic radiation pattern \cite{iliadis} and no corrections for angular correlation effects were necessary. For each of these templates, $2\times 10^6$ decays were simulated to ensure sufficient statistics. The simulated events were then sorted with the same timing and energy gates as the experimental data.
\par
\begin{figure}
    \centering
    \includegraphics[width=.40\textwidth]{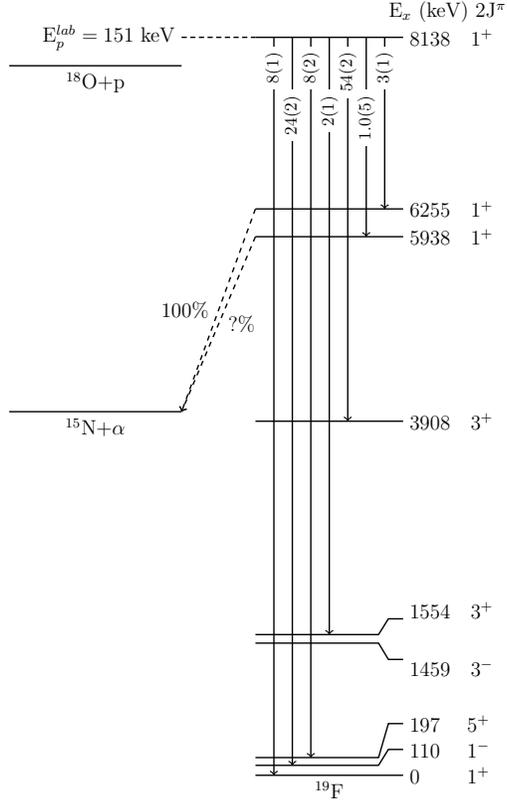}
    \caption{Energy level diagram of \atom{19}{F}. Only the relevant levels are shown. Primary $\gamma$-ray transitions are shown as solid vertical arrows. Secondary transitions have been omitted. Excited state energies and branching ratios as reported in Ref.~\cite{Tilley}. The 6255-keV excited state decays to \atom{15}{N} via $\alpha$-emission with a branching ratio of $100\%$. The 5938-keV state may also decay this way; however, the branching ratio is yet undetermined.}
    \label{fig:oxydecayscheme}
\end{figure}
\par
For the $5938$-keV and $6255$-keV excited states, $\gamma$-ray decay competes with $\alpha$-particle emission to \atom{15}{N} (S$_\alpha=-4013.80$ keV) \cite{wang}. For the $6255$-keV state, $\alpha$-particle emission is the only decay channel \cite{Tilley}, indicating that the R$\rightarrow6255$ primary transition does not give rise to $\gamma\gamma$-coincidences. For this reason, we exclude the R$\rightarrow6255$ template from the fit of the coincidence spectrum. For the $5938$-keV level, the width of the $\alpha$-particle decay channel has not yet been reported in the literature. Transitions to lower lying states in $^{19}$F have been reported in Ref.~\cite{rogers}; however, full-energy peaks for these secondary transitions are absent from our singles spectrum while no evidence of the R$\rightarrow5938$ transition (primary or secondary) exists in the coincidence spectrum. 
To generate templates for this transition, we assumed an $\alpha$-particle decay branching ratio of $90\%$ in the simulations. 
This value was chosen for being consistent with the poor $\gamma\gamma$-coincidence efficiency observed, while also including the reported $\gamma$-ray branch.
\par
To describe the component of the singles and coincidence spectra attributable to environmental radionuclides, background runs were recorded before and after the resonance run. Background was measured for a total of 20 hours. The combined spectrum, sorted with and without coincidence conditions, was then used as a template in the analysis of the singles and coincidence spectrum, respectively. Beam-induced background was not identified in the measured spectrum and was therefore disregarded. An unidentified  peak was observed in the singles spectrum at E$_{\gamma}= 2801.5 \pm 2.7$ keV, having an efficiency-corrected intensity of $3400\pm500$ counts. A template was made to account for this peak by simulating a mono-energetic $\gamma$-ray source of the same energy; however, lacking evidence to substantiate this peak as a new transition (primary or secondary), it was considered background in the analysis.   
\par

\begin{table*}   
\centering  
\begin{threeparttable}
  
    \caption{Present results for the number of reactions (N$_R^{partial}$) and primary branching ratios (BR) for the E$_p^{lab}=151$ keV resonance\tnote{f} in \oxy{}.}
    \par\medskip
    \begin{tabular}{ l | c c  |  c  c  | c }
    \hline \hline
     & \multicolumn{2}{c}{N$_R^{partial}$} & \multicolumn{3}{c}{ BR ($\%$)}\\ \hline
    Transition & singles\tnote{b}  & coincidence\tnote{b} & singles & coincidence\tnote{e} & \multicolumn{1}{c}{Tilley \textit{et al.}~\cite{Tilley}}\\ 
    R$\rightarrow0 $  & 23000(1300)  & 18000(2000)   & 8.5(5)  & 6.8(8)\tnote{d} & 8(1) \\
    R$\rightarrow110 $& 64000(2000)   & 65000(3000) &  23.5(6)& 24.7(10) & 24(2) \\
    R$\rightarrow197 $ &19500(1300)    & 21000(2000) &  7.1(5) & 7.9(9)  & 8(1)\\
    R$\rightarrow1554 $ &3400(600)  & 2700(600)      & 1.2(2)  & 1.0(2)  & 2(1)\\
    R$\rightarrow3908 $ &158000(1000) & 153000(1000) & 57.4(5) & 58.0(6)& 54(2)\\
    R$\rightarrow5938 $ &2400(500)   &  $<3400$   & 0.9(2)  &  $<1.3$   & 1.0(5)\\
    R$\rightarrow6255 $ &3900(500)   & \tnote{c}  & 1.4(2)  & \tnote{c}   & 3(1)\\
    \hline 
    N$_R^{\text{total}}$ & 274000(14000)\tnote{a}   & 264000(13000)\tnote{a}   &  &     &   \\
    \hline \hline
    \end{tabular}
\begin{tablenotes}
\item [a] \footnotesize Systematic uncertainties added in quadrature (see text). 
\item [b] \footnotesize Uncertainties correspond to the $68\%$ credible interval (95\% for upper-limits) of posterior distribution.
\item [c] \footnotesize Transition was unobserved in the coincidence spectrum because of $\alpha$-particle decay and is excluded from the coincidence fit. The partial number of reactions measured in the singles fit is used to calculate N$_R^{\text{total}}$ in the coincidence analysis.
\item [d] \footnotesize Coincidence analysis of the ground state transition based entirely on escape peaks and Compton continuum.
\item [e] \footnotesize Branching ratios sum to $98.4\%$. The missing $1.6\%$ is attributed to the R$\rightarrow5938$ and R$\rightarrow 6255$ transitions, which are unobserved in coincidence.
\item [f] \footnotesize Accumulated charge was $38$ mC, with a beam intensity of $\approx 30$ $\mu$A.

\end{tablenotes}
     \label{tab:oxyresults}
   \end{threeparttable}
\end{table*} 
The singles and coincidence spectra were fit using the method described in Sec.~\ref{sec:analysis}. For these fits, the analysis was limited to energies above a low-energy threshold (200 keV and 800 keV for singles and coincidence, respectively).  
From the fit, the estimated contributions of each template to the experimental pulse height spectrum were obtained; they are plotted (in shades of blue/purple) alongside the experimental pulse height spectrum (black) in Fig.~\ref{fig:pulseheightspectrum}. The sum of all the templates (red) agrees with the experimental data, corroborating the validity of the fit results. 
\par
The partial number of \atom{18}{O}$+ \text{p}$ reactions was calculated for each primary transition using the obtained posterior probability distributions. Primary branching ratios are calculated using Eq.~\ref{eq:br}; the results are listed in Table~\ref{tab:oxyresults}.  We find that both the singles and coincidence fits yield $\gamma$-ray branching ratios that are in agreement with each other. 
The discrepancy ($\approx$ 1.7\%) in the ground state (R$\rightarrow$0) branching ratio can be explained by the absence of a full-energy ground state peak in the coincidence spectrum. Consequently, the ground state branching ratio extracted from the coincidence spectrum is entirely based on the intensities of the escape peaks and the Compton continuum caused by the ground state transition.
The R$\rightarrow5938$ branch was unobserved in our coincidence spectrum. However, the upper-limit reported in the coincidence analysis is consistent with the intensity and branching ratio obtained from the singles analysis. This suggests that our estimate for the $\alpha$-decay branching ratio is compatible with observations, though more data is required to resolve this ambiguity. 
Our $\gamma$-ray branching ratios are also in agreement with the values from Ref.~\cite{Tilley}. However, we have reduced the uncertainties by a factor of $4$.
\par
The resonance strength can be calculated from the experimental thick-target yield, according to \cite{iliadis}:
\begin{equation} \label{eq:strength}
\omega\gamma = \frac{2 \epsilon_{\text{eff}}}{\lambda_r^2} \times  \frac{N_R^\text{total}}{\mathcal{N}_p}
\end{equation}
where $\mathcal{N}_p$ is the number of incident protons, $\lambda_r$ is the de Broglie wavelength of the incident proton, and $\epsilon_{\text{eff}}$ is the effective stopping power, derived from Bragg's rule. Assuming a target stoichiometry of \atom{}{Ta}$_2$\atom{18}{O}$_5$, the effective stopping power can be obtained from SRIM \cite{SRIM}. The total resonance strength error is obtained by adding the statistical uncertainty (i.e., from the decomposition of the measured spectrum) and systematic uncertainties, e.g., current integration (3$\%$), effective stopping power (4$\%$), and simulated detector response (5$\%$). The statistical uncertainty of the fit amounted to $0.5\%$ and $0.7\%$ for singles and coincidence, respectively.
\par
The measured $\omega\gamma$ values from the singles and coincidence data, along with literature values, are listed in Table ~\ref{tab:allwg}. It can be seen that the resonance strength values derived from the singles and coincidence fits are in agreement. Our mean value, $\omega\gamma_{pres}=1.05\pm0.08$ meV, also agrees with the results of Becker \textit{et al.} \cite{becker}, Wiescher \textit{et al.} \cite{wiescher}, and Vogelaar \textit{et al.} \cite{vogelaar}.

\begin{table*}
    \centering
\begin{threeparttable}
    \caption{Resonance strengths, $\omega\gamma$, measured in the present work and a comparison to literature values ($\omega\gamma$ in units of meV). }\par\medskip
        \begin{tabular}{ l  r  }
        \hline \hline
        \multicolumn{2}{l}{E$_p^{lab}=151$ keV, \oxy{}} \\
        \hline
        \textbf{present(sing)}  &    $1.07\pm0.08$ \\
        \textbf{present(coin)}  &     $1.03\pm0.07$ \\
        \textbf{present(mean)}\tnote{a}  &  $1.05\pm0.08$ \\

        Vogelaar \cite{vogelaar}  &  $0.92\pm0.06$ \\
        Becker \cite{becker}  &  $1.1\pm0.1$ \\
        Wiescher \cite{wiescher}  &  $1.0\pm0.1$ \\
     
        \rule{0pt}{0.2ex}  & \\
  \hline     
   \multicolumn{2}{l}{E$_p^{lab}= 317$ keV, \mg{}}\\
        \hline
        \textbf{present(sing)}  & $28.7\pm3.3$  \\
        \textbf{present(coin)}   & $30.5\pm3.5$ \\
        \textbf{present(mean)}\tnote{a}  & $29.6\pm3.4$ \\

        Limata \cite{Limata}      &  $30.7 \pm 1.7$   \\
        Iliadis \cite{iliadis90}  &  $30 \pm 4$   \\
        Endt \cite{endt86}        & $36 \pm 4$ \\
        Anderson \cite{anderson80}& $31 \pm 4$ \\
        Elix \cite{elix79}        & $ 24 \pm 6$ \\
        \hline \hline
        \end{tabular}
    \begin{tablenotes}
\item[a] \footnotesize  Mean value from singles and coincidence data of present work.

\end{tablenotes}
    \label{tab:allwg}
\end{threeparttable}
\end{table*}

\section{\mg} \label{sec:25mg}
We will now apply our method to the $\text{E}_p^{lab}=317$ keV and $\text{E}_p^{lab}=435$ keV resonances in \mg{}, which have complex decay schemes consisting of 25 and 11 primary transitions, respectively, each with a multitude of subsequent secondary decays \cite{Limata,endt88}.
Our goals are to determine the branching ratios via the measurement of partial reaction numbers and to measure the strength of the 317-keV resonance.
To that end, thick-target excitation functions were recorded at both resonances.
These are shown in Fig.~\ref{fig:yields}.
A Markov chain Monte Carlo method (separate from that of Sec. \ref{sec:analysis}) was used to obtain a fit (red line) and determine the target thicknesses in energy units, $\Delta$E$_{317}=12.8\pm0.4$ keV and $\Delta$E$_{435}=12.2\pm 0.4$ keV. 
The \mg{} reaction was then measured at each resonance at energies corresponding to their maximum yield heights. We accumulated $6400 ~\mu$C and $1807 ~\mu$C of charge at the $317$-keV and $435$-keV resonances, respectively. The resonance data were measured and then sorted into singles and coincidence spectra, where the condition of Eq.~\ref{eq:2} was used for the coincidence energy gates. 
\begin{figure}
    \centering
    \includegraphics[width=.40\textwidth]{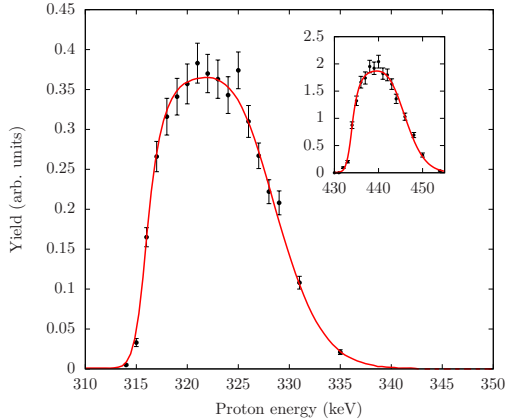}
    \caption{(Color Online) Thick-target excitation functions for the E$_p^{lab}=317$ keV and  E$_p^{lab}=435$ keV resonances in \mg{}. The red line represents a fit used to extract the target thickness.}
\label{fig:yields}
\end{figure}
\par
The decay schemes of these resonances are shown in Fig.~\ref{fig:mgdecay}. For reasons of clarity, only the strongest primary transitions for the $\text{E}_p^{lab}=317$ keV resonance are included. Templates were generated for each primary transition by simulating $2\times10^6$ decays. Excited state energies and secondary transition branching ratios used in the simulations were adopted from Ref.~\cite{endt88}. 
The isomeric state at E$_x= 228$ keV in \atom{26}{Al} beta decays to the \atom{26}{Mg} ground state with a half-life $\tau_{1/2} = 6.3465(8)$ s \cite{nubase12}. This decay proceeds via positron emission with an end-point energy E $ = 3210.45(6)$ keV, producing Bremsstrahlung and 511-keV annihilation radiation. 
We took this into account by simulating the emission of a positron on the surface of the target
\footnote{The effect of the MgO layer (not simulated) is negligible. Consider a $1.5$ MeV positron incident on $45$ $\mu$g/cm$^2$ MgO. 
According to the ESTAR database \cite{NIST}, the total energy loss due to collisional and radiative effects is only $70$ eV, less than $0.01\%$ of the incident energy.
} 
and including this template in the analysis of the singles spectra. 
\par
Angular correlation effects for the strongest primary transitions, where $\Delta J =0,\pm1$ and $\pi_f=-\pi_i$, were estimated using the formalism in Ref.~\cite{iliadis}. 
Each of these transitions was assumed to proceed via electric-dipole radiation (E1), with no angular momentum mixing. The channel-spin mixing ratios for the primary transitions are unknown. However, anisotropies for the strongest transitions in \mg{} were measured by Ref.~\cite{deneijs} and suggests that the $317$-keV resonance proceeds mainly with channel-spin $j_s=3$. Thus, the theoretical angular correlation functions for the $317$-keV resonance are given by
\begin{eqnarray}  
W(\theta)_{R\rightarrow4^+}&=&1+0.125\ P_2(\cos\theta), \\
W(\theta)_{R\rightarrow3^+}&=&1-0.375\ P_2(\cos\theta),  \\
W(\theta)_{R\rightarrow2^+}&=&1+0.3\ P_2(\cos\theta). 
\end{eqnarray}
The theoretical angular correlation functions for the $435$-keV resonance are insensitive to the channel-spin and are given by
\begin{eqnarray}  
W(\theta)_{R\rightarrow5^+}&=&1-0.10\ P_2(\cos\theta),\\
W(\theta)_{R\rightarrow4^+}&=&1-0.20\ P_2(\cos\theta),\\
W(\theta)_{R\rightarrow3^+}&=&1+0.275\ P_2(\cos\theta).
\end{eqnarray}
The angular correlation effects were incorporated into the simulations of the template spectra, as described in Sec.~\ref{sec:corrections}. 
\par
We accounted for environmental contaminants by measuring the room background for a total of 340 hours (see Sec.~\ref{sec:bkgd}). We also observed peaks originating from beam-induced backgrounds from, for example, \flourine{} and \carbon{}. We simulated templates for each of these reactions, and included them in the analysis along with the templates for the \atom{25}{Mg}+p reaction.
\par
\begin{figure}
    \centering
    \includegraphics[width=.40\textwidth]{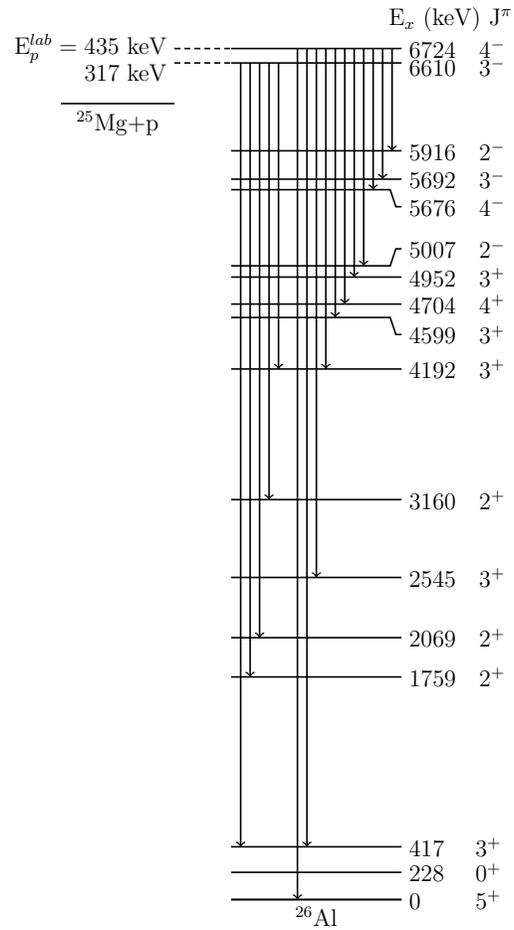}
    \caption{Excited state energies with primary transitions as reported in Ref.~\cite{endt88}. Secondary transitions have been omitted. For the $E_p^{lab}=317$ keV cascade, branches weaker than $5\%$ have also been omitted for reasons of clarity.}
\label{fig:mgdecay}
\end{figure}
\par
\begin{figure}
    \centering
    \includegraphics[width=.6\textwidth]{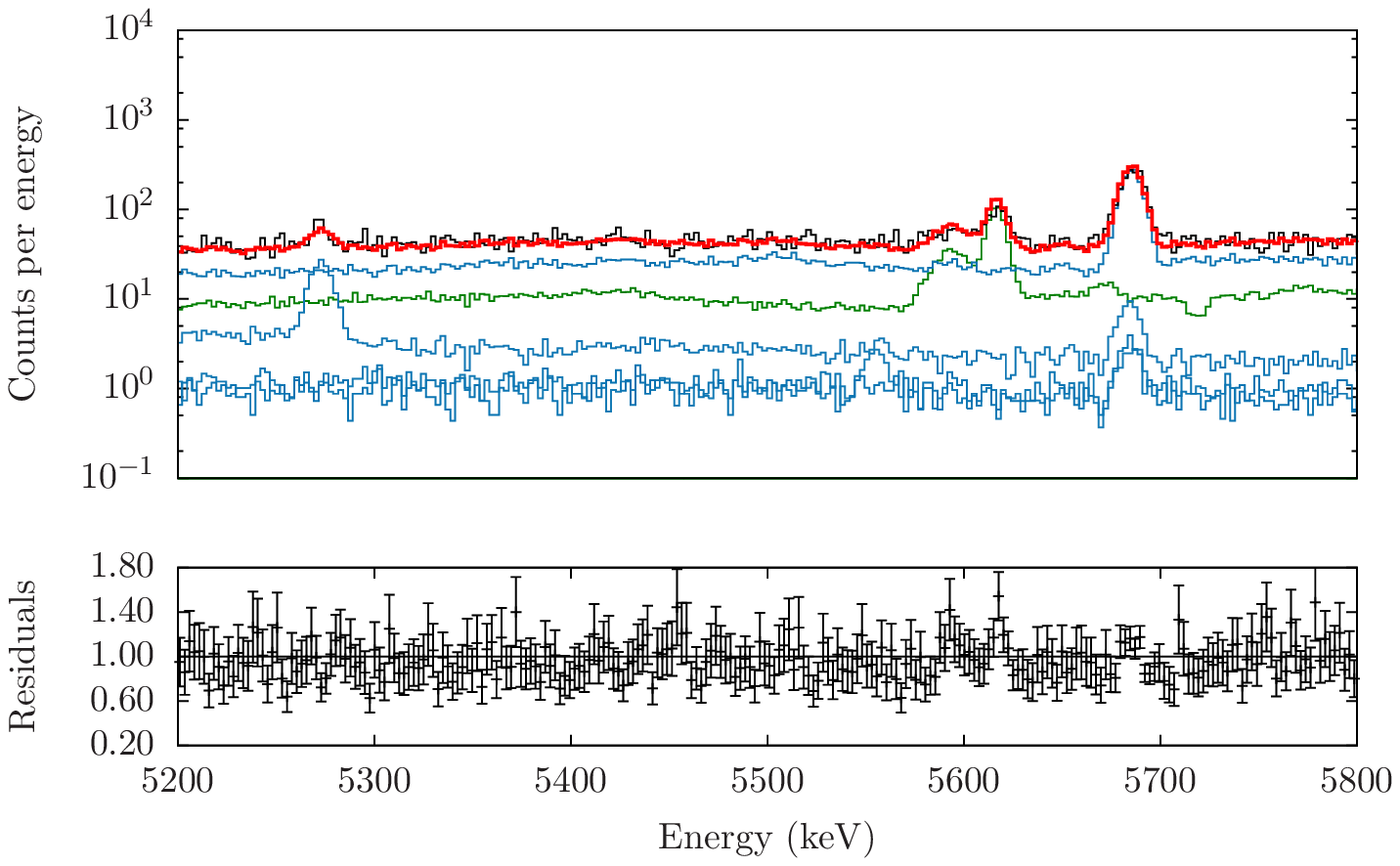}
\caption{(Color online) Top: a segment of the measured pulse height spectrum (black) obtained for the E$_r^{lab}$ = 317 keV resonance in \mg{}. 
The contributions of the different primary decays (i.e., scaled templates), including all associated secondary transitions, are shown in shades of blue/purple. The red line represents the sum of all the templates. \flourine{} background is represented by the green line. Bottom: the residuals for the fit, calculated by taking the ratio of the predicted counts, i.e., the template sum, over the observed counts.}
\label{fig:residuals}
\end{figure}
\par
The measured spectra for both resonances were fit using the method outlined in Sec.~\ref{sec:analysis}, where the lower fit limit was set to 400 keV and 700 keV, respectively. The decomposed 317-keV spectrum is shown in Fig.~\ref{fig:residuals}. The R$\rightarrow417$ single escape peak (rightmost), can be seen with the \flourine{} background single escape peak (second-to-right). A sum-peak arising from the R$\rightarrow1759$ transition is also visible (leftmost). Below, the residuals are plotted, indicating that the predicted spectrum sufficiently reproduces the observed data.
Posterior probability distributions for the R$\rightarrow0$ and R$\rightarrow5676$ transitions are shown in Fig.~\ref{fig:posteriors}. The credible intervals, shown in red, denote the highest probability region for the partial reaction number.
For R$\rightarrow$0 (top), the distribution is sharp, indicating that the transition has strong features in the experimental spectrum. Conversely, the R$\rightarrow5676$ transition (bottom), is heavily skewed towards zero, suggesting little or no spectroscopic evidence of the transition.   
\par
The measured partial number of reactions and $\gamma$-ray branching ratios are listed in Table~\ref{tab:mgresultstable435} (E$_r^{lab}$=435 keV) and  Table~\ref{tab:mgresultstable} (E$_r^{lab}$=317 keV). For completeness, we also performed an analysis of the $435$-keV singles data using the more simple maximum-likelihood estimate. For strong transitions, we find that the maximum-likelihood method yields results identical to their Bayesian counterpart. The distinction between the two methods is made more apparent by the weak transitions. Consider the R$\rightarrow4952$ and R$\rightarrow5692$ transitions, for which the maximum-likelihood yielded $100\pm300$ and $300\pm600$ reactions, respectively. The uncertainty for each of these, calculated using the profile likelihood~\cite{james}, suffer from over-coverage. That is, the confidence interval extends beyond physically allowed regions (ie., N$_{R}^{partial} < 0$). This inconsistency is absent from the Bayesian derived results, which allow for statistically robust upper-limits without deferring to more specialized interval construction methods. 
\par
It can be seen that the $\gamma$-ray branching ratios and total number of reactions obtained using the singles and coincidence Bayesian analysis are in agreement. 
This is especially compelling considering the $435$-keV resonance, where the ground-state transition branching ratio measured in coincidence is in agreement with the singles result, despite the absence of the full-energy peak in the coincidence spectrum. 
Overall, the measured branching ratios are in agreement with Endt \textit{et al.}~\cite{endt88} and Limata \textit{et al.}~\cite{Limata}. We would like to emphasize that Endt \textit{et al.} and Limata \textit{et al.} accumulated about $3.5$ C and $7.5$ C in their measurement of the E$_r^{lab}$=435 keV and 317 keV resonances. These values are about 3 orders of magnitude larger than the charges accumulated in the present work.
We also report a ground-state feeding fraction, $f_0$, for both resonances. 
This is obtained by measuring the fraction of events, $1-f_0$, which instead decay to the isomer state.
Using the partial number of reactions measured for the isomer decay, the ground-state feeding fraction is determined to be $88.1\pm 0.3\%$ and $97.8\pm 0.3\%$ for the $317$-keV and $435$-keV resonances, respectively. These are in agreement with the measurements in Limata \textit{et al.} ($87.8\pm1.0\%$) and Ref.~\cite{endtrolfs} ($96\pm1\%$).
\par
\begin{figure}
    \centering
    \includegraphics[width=.6\textwidth]{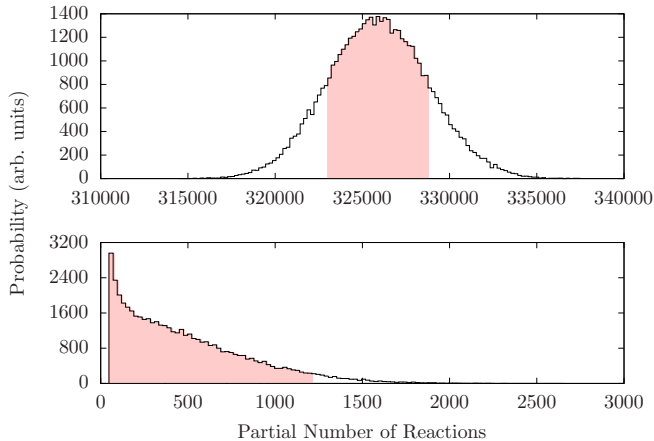}
\caption{(Color online) Posterior probability distributions for transitions in the E$_r^{lab}=435$ keV cascade (singles). Top: The $68\%$ credible interval for the R$\rightarrow$ 0 transition (shown in red).  Bottom: The $95\%$ credible interval for the R$\rightarrow$ 5676 transition, used to define the upper-limit. }
\label{fig:posteriors}
\end{figure}
\sethlcolor{yellow}
\par
The resonance strength of the $317$-keV resonance was determined using Eq.~\ref{eq:strength}. The effective stopping power is given by
\begin{equation} \label{eq:eff}
\epsilon_{\text{eff,$x$}} = \frac{\text{M}_{\text{Mg}}}{\text{M}_{\text{Mg}}+\text{M}_{\text{p}}}\bigg[ \epsilon_{\text{Mg,$x$}} + \frac{n_{\text{O}}}{n_{\text{Mg}}} \epsilon_{\text{O,$x$}}\bigg]
\end{equation} \\
where M$_{\text{p}}$ and M$_{\text{Mg}}$ are the mass of the proton and the \atom{25}{Mg} atom, $\epsilon_{\text{Mg,$x$}}$ and $\epsilon_{\text{O,$x$}}$ are the stopping powers of protons at bombarding energy E$_p^{lab}=x$  (supplied by SRIM \cite{SRIM}), and $n_{O}/n_{Mg}$ is the target stoichiometry. Unfortunately, the stoichiometry of the \atom{25}{Mg} target is largely unpredictable, owing to the evaporation process.
However, Eq.~\ref{eq:eff} holds for both bombarding energies used in this experiment. We can exploit this to obtain the target stoichiometry by first measuring $\epsilon_{\text{eff,435}}$ (via Eq.~\ref{eq:strength}) using the average maximum yield (singles and coincidence) at the $435$-keV resonance ($\omega\gamma=9.42\pm0.65 \times 10^{-2} $ eV \cite{powell}). This yields a target stoichiometry of $n_{O}/n_{Mg} = 0.61 \pm 0.18$, which can then be used to calculate the resonance strength, $\omega\gamma(317 \text{ keV})$.  
Uncertainties were derived by adding systematic and statistical errors in quadrature. The systematic contributions originate from the current integrator ($3\%$), the simulated detector response ($5\%$), and effective stopping power ($10\%$). The statistical uncertainty of the fit amounted to $0.7\%$ and $1.1\%$ for singles and coincidence, respectively. 
\par
The results of the present measurement are listed in Table \ref{tab:allwg} alongside literature values. We find again that the resonance strengths obtained using the singles and coincidence fit are in agreement with each other. This corroborates our claim that the $\gamma\gamma$-spectrometer (described in Sec.~$\ref{sec:detectors}$) has been modeled to sufficient accuracy. Our mean value amounts to $\omega\gamma_{pres}=29.6 \pm 3.4$ meV, which agrees with the previously reported values. The present measurement has an uncertainty comparable to that of the strengths reported in Refs.~\cite{iliadis90,endt86,anderson80,elix79}. 
\par
\newpage
\begin{table*}[h!]
    \centering
    \begin{threeparttable}
    \caption{Present results for the partial number of reactions (N$_R^{partial}$) and primary branching ratios (BR) for the E$_p^{lab}=435$ keV resonance\tnote{d} in \mg{}.}\par\medskip
    \begin{tabular}{ l | c  c | c c  | c }
    \hline \hline
     & \multicolumn{2}{c}{N$_R^{partial}$} & \multicolumn{3}{c}{ BR $\%$}\\ \hline 
    Transition & singles\tnote{b}  & coincidence\tnote{b}  & singles & coincidence & \multicolumn{1}{c}{Endt \textit{et al.} \cite{endt88}\tnote{e}}  \\ 
R$\rightarrow0    $    & 326000(3000)  & 325000(5000)     & 49.7(4)  & 51.2(5)& 50(1) \\
R$\rightarrow417  $    & 194000(2000)  & 173000(4000)  & 29.6(4)  & 27.3(5) &  29.9(9) \\
R$\rightarrow2545 $    & 17000(1000)      & 18300(1300)   & 2.6(2)   & 2.9(2)&  1.60(7)\\
R$\rightarrow4192 $    & 51000(1000)    & 50300(1500)     & 7.8(3)   & 7.9(3) & 6.4(2)\\
R$\rightarrow4599 $    & 37000(1000)    & 38200(1400)     & 5.7(2)   & 6.0(2) & 5.3(2)\\
R$\rightarrow4705 $    & 28000(1000)    & 28000(1000)     & 4.3(2)   & 4.4(2) & 4.1(1)\\
R$\rightarrow4952 $\tnote{c}    & $< 630$  & $ < 720 $  & $ < 0.10$            &  $ < 0.11 $            & 0.064(7)  \\
R$\rightarrow5007 $\tnote{c}    & \ $< 2300$  &\  $ < 1100 $  & $ < 0.36 $ & $ < 0.18 $  & 0.017(4)  \\
R$\rightarrow5676 $\tnote{c}    & \ $< 1200$  & $ < 970$  & $ < 0.18$        & $ < 0.15 $         & 0.034(4)  \\
R$\rightarrow5692 $\tnote{c}     & $< 960$  & \ \ $ < 1600 $  & $ < 0.14$     & $ < 0.25 $      & 0.019(4)  \\
R$\rightarrow5916 $\tnote{c}   & \ \ $< 1100$  & \ \ $ <1800 $  & $ < 0.16$ & $ < 0.28 $  & 0.024(4)  \\
$228\xrightarrow{\beta^+}\,^{26}\text{Mg}$    & 14500(1700) \tnote{f} &   &  &  &  \\
    \hline
    N$_R^{\text{total}}$ & 655000(33000)\tnote{a}   & 634000(33000)\tnote{a}   &   &    &  \\
    \hline \hline
    \end{tabular}
    \begin{tablenotes}
    \item [a] \footnotesize Systematic uncertainties added in quadrature. 
    \item [b] \footnotesize Uncertainties correspond to the $68\%$ credible interval ($95\%$ for upper-limits) of posterior distribution.
    \item [c] \footnotesize Full-energy peak was unobserved.
    \item [d] \footnotesize Accumulated charge was $1807$ $\mu$C, with a beam intensity of $\approx 6$ $\mu$A.
    \item [e] \footnotesize Accumulated charge was $3.5$ C.
    \item [f] \footnotesize Decay of the 228-keV isomer state to the $^{26}$Mg ground state. Intensity is not included in N$_R^{\text{total}}$.
    \end{tablenotes}
    \label{tab:mgresultstable435}
\end{threeparttable}
\end{table*} 

\begin{table*}[!]     
  \centering
  \begin{threeparttable}
  \caption{ Present results for the partial number of reactions (N$_R^{partial}$) and primary branching ratios (BR) for the E$_p^{lab}=317$ keV resonance\tnote{d} in \mg{}.}
    \begin{tabular}{ l | c  c |  c c | c }
    \hline \hline
     & \multicolumn{2}{c}{N$_R^{partial}$} & \multicolumn{3}{c}{ BR $\%$}\\ \hline 
    Transition & singles\tnote{b} & coincidence\tnote{b}& singles & coincidence & \multicolumn{1}{c}{Limata \textit{et al.} \cite{Limata}\tnote{e}}\\
R$\rightarrow0 $\tnote{c} & $5400(1200)$      & $6000(2000)$  & $0.65(14)$  & $0.7(2)$ & 0.058(4) \\
R$\rightarrow417 $  & $262000(3000)$  & $285000(5000)$        & 31.7(3)  & 32.4(5)& 31.8(5) \\
R$\rightarrow1759$ & $124000(2000)$  & $133000(2000)$         & 15.1(2)  & 15.2(2) &  16.1(3)\\
R$\rightarrow2069 $       & $44600(1400)$   & $47000(2000)$   & 5.4(2)   & 5.3(2)& 6.0(1)\\
R$\rightarrow2365 $\tnote{c} & $6000(1300)$   & $< 4500$      & $0.73(16)$  & $<0.51$ & 0.47(2) \\
R$\rightarrow2545 $       & $16500(1300)$   & $15000(2000)$   & 1.99(16)   & 1.7(2) & 1.46(3)\\
R$\rightarrow2661 $        & $8000(1000) $  & $5800(1200)$    & 1.0(1)   & 0.66(14) & 1.00(2)\\
R$\rightarrow2913 $       & $23000(1000)$   & $21400(1400)$   & 2.79(14)   & 2.43(16) & 3.04(5)\\
R$\rightarrow3073 $\tnote{c} & $<1400$   & $ < 930$           & $<0.16$  & $<0.11$ & 0.11(4) \\
R$\rightarrow3160 $ & $92700(1700)$   & $94000(2000)$         & 11.2(2)  & 10.7(2) & 11.4(2)\\
R$\rightarrow3596 $        & $37800(1400)$   & $43000(2000)$  & 4.57(16)   & 4.9(2) & 4.29(7)\\
R$\rightarrow3675 $\tnote{c} & $<4700$   & $<4500$            & $<0.56$ & $<0.51$ & 0.86(13) \\
R$\rightarrow3681 $      & $13000(1000)$   & $15000(1500)$    & 1.56(14)   & 1.7(2) & 1.09(3)\\
R$\rightarrow3750 $        & $9000(1000)$   & $10000(1000)$   & 1.1(1)   & 1.1(1) & 0.92(2)\\
R$\rightarrow3963 $\tnote{c} & $4000(1000)$   & $5000(1000)$  & $0.46(13)$  & $0.55(14)$ & 0.17(1)\\
R$\rightarrow4192 $       & $146000(2000)$  & $160000(3000)$  & 17.7(3)  & 18.1(3) & 19.1(3)\\
R$\rightarrow4206 $\tnote{c} & $<1700$  & $<2000$             & $<0.21$  & $<0.23$ & 0.25(2) \\
R$\rightarrow4349 $\tnote{c} & $2400(1000)$   & $4000(1000)$  & $0.3(1)$  & $0.5(1)$ & 0.03(1) \\
R$\rightarrow4548 $       & $15500(1300)$   & $17600(1700)$   & 1.88(15)  & 2.0(2) & 1.30(1)\\
R$\rightarrow4599$\tnote{c} & $ < 3400 $   & $3000(1400)$     & $<0.41$   & $0.34(16)$ & 0.12(1) \\
R$\rightarrow4622$\tnote{c} & $ 2600(1000)$   & $2700(1100)$  & $0.3(1)$  & $0.30(13)$ & 0.28(7) \\
R$\rightarrow4940 $\tnote{c} & $ 2400(800)$   & $<3400$       & $0.3(1)$  & $<0.39$ & 0.08(1) \\
R$\rightarrow5396 $\tnote{c} & $ 3800(900)$   & $4000(1000)$  & $0.5(1)$  & $0.44(13)$ & 0.22(2)\\
R$\rightarrow5726 $\tnote{c} & $ 2400(800)$   & $<2000$       & $0.3(1)$  & $<0.23$ & 0.10(1)\\
R$\rightarrow5916 $\tnote{c} & $ < 1400$   & $<6000$          & $<0.16$   & $<0.69$ & 0.09(2)\\
$228\xrightarrow{\beta^+}\,^{26}\text{Mg} $    & 102000(2000)\tnote{f} &   &  &  &  \\
    \hline 
    N$_R^{\text{total}}$ & 826000(42000)\tnote{a}   & 879000(45000)\tnote{a}   &   &   &  \\
    \hline \hline
    \end{tabular}
    \begin{tablenotes}
\item [a] \footnotesize Systematic uncertainties added in quadrature.
\item [b] \footnotesize Uncertainties correspond to the $68\%$ credible interval ($95\%$ for upper-limits) of posterior distribution.
\item [c] \footnotesize Full-energy peak was unobserved.
\item [d] \footnotesize Accumulated charge was $6400 \mu$C, with a beam intensity of $\approx 12 \mu$A.
\item [e] \footnotesize Accumulated charge was $7.5$ C.
\item [f] \footnotesize Decay of the 228-keV isomer state to the $^{26}$Mg ground state. Intensity is not included in N$_R^{\text{total}}$.
   \end{tablenotes}
    \label{tab:mgresultstable}
    \end{threeparttable}
\end{table*} 
\newpage
\section{Conclusion}
\label{sec:conclusion}
In this work, we presented a novel data-analysis method for $\gamma$-ray spectroscopy. Using this method, the primary transition $\gamma$-ray branching ratios and total number of reactions can be measured, not only from full-energy peaks, but from the entire spectrum. This approach is advantageous in the analysis of complex $\gamma$-ray spectra, where full-energy peaks are often obscured by, for example, escape peaks, environmental background, or beam-induced background. We also provided a strategy for simulating the template spectra using Monte Carlo techniques and recommended procedures for correcting templates for experimental artifacts such as Doppler shifts, detector resolution, and angular correlation effects.
\par
We then applied our analysis to resonances in \atom{18}{O}+p and \atom{25}{Mg}+p. Measured $\gamma$-ray pulse height spectra were decomposed into separate components, each corresponding to a given primary transition plus the subsequent secondary decays.  The contribution of each transition to the experimental data was determined using a Bayesian approach, where the finite statistics of the recorded data and the simulated template spectra were implicitly considered through our choice of the likelihood function. We found for the $151$-keV resonance in \atom{18}{O}+p that our measured primary transition branching ratios were more precise than the literature values. We also found that the resonance strength determined in the present work is in agreement with previous measurements. With regard to the \atom{25}{Mg}+p reaction, we measured the branching ratios and ground-state feeding fractions of the $317$-keV and $435$-keV resonances and found that they, too, were in overall agreement with the literature values. Finally, the strength of the $317$-keV resonance was measured and found to be competitive with many previous measurements, despite the poor statistics collected in the present work. These results demonstrate that our method yields results consistent with the most sensitive previous measurements. 
\section*{Acknowledgments}
This work was supported by the U.S. Department of Energy under Contract no. DE-FG02-97ER41041. The authors would like to express their gratitude to the reviewers for their thoughtful suggestions. Comments from Richard Longland were also greatly appreciated.
\newpage
\section*{Appendix}
\label{sec:appendix}
The fundamental difference between Bayesian \cite{bayes}  and classical (frequentist) statistics can best be understood by examining interval construction in both fields. For the frequentist, confidence intervals are characterized by the concept of \textit{coverage}, which seeks to answer: `If this experiment were to be repeated and reanalyzed $N$ times, what fraction of the new confidence intervals contains the (fixed) parameter value?' This is a natural question for a frequentist, as they maintain that data are a repeatable random sample, with all parameters being fixed. The Bayesian, on the other hand, has no such concept of coverage. Bayesian inference is fully conditioned on the observed data, while parameters values are unknown and described by probability distributions. Thus, the confidence interval associated with Bayesian inference, the \textit{credible} interval, quantifies the belief that the parameter value lies within the interval, given the data  \cite{james}. 
\par
To analyze data using Bayesian inference, we assign distributions to describe model parameters, $\bm{\theta}$: both the joint unconditional probability distribution function (PDF), $P(\bm{\theta})$, and the joint conditional PDF, $P(\bm{\theta}|\bm{D})$, are taken to represent the \textit{degree of belief} in different values of $\bm{\theta}$. For the joint conditional distribution, the degree of belief is conditional on the data set, $\bm{D}$. These are known as the prior and posterior PDF, respectively. The prior distribution summarizes the state of knowledge before performing a measurement. It can be either \textit{informative}, in the sense that it has been influenced by past experiments or theory, or \textit{non-informative}, i.e., any value of $\bm{\theta}$ is equally likely. To calculate the posterior distribution, practitioners ``update'' their prior beliefs using Bayes' theorem:
\begin{equation}
\label{eq:a1}
  P(\bm{\theta} |\bm{D}) = \frac{ P( \bm{D} |\bm{\theta} )P(\bm{\theta} )    }{\int_{\theta}P(\bm{D}|\bm{\theta})P(\bm{\theta})} \;\;\;,
\end{equation}
where $P( \bm{D} | \bm{\theta} )$ represents the likelihood of the collected data, $\bm{D}$, given the parameter values, $\bm\theta$. To obtain a posterior probability distribution for a single parameter, the parameter is marginalized out:
\begin{equation}
\label{eq:a2}
  P(\theta_0 | \bm{D}) =\int_{\theta_1,\theta_2, \ldots,\theta_k } P(\bm{\theta}|\bm{D})P( \bm{\theta}) \;\;\;.
\end{equation} 
The marginalized posterior distribution, $P(\theta_0|\bm D)$, summarizes all one's knowledge or belief concerning $\theta_0$, given both the prior belief and the experimental data, $\bm D$. All the usual statistically meaningful quantities can be obtained from this function, e.g., the location and spread of the distribution.
A credible interval, $[\theta_L,\theta^U]$, with probability content $\beta$, can be defined by:
\begin{equation}
\int_{\theta_L}^{\theta^U} P(\theta_0 |\bm  D) d\theta_0 = \beta \;\;\; .
\end{equation} 
By this definition, the interval $[\theta_L,\theta^U]$ contains a fraction $\beta$ of one's total belief about $\theta$. Frequently, the narrowest interval is chosen, corresponding to the region of highest probability density. In the present work, upper-limits are used to summarize the posterior distributions of weak transitions. We can define an upper-limit by taking $\theta_L \rightarrow 0$. In this case, the interval suggests that the parameter value is less than the upper-limit value, $\theta^U$, at the $\beta$ credibility level. 

\section*{References}
\bibliography{manuscript_review}
\end{document}